\documentclass[aps,pre,twocolumn,groupedaddress,showpacs]{revtex4-1}
\usepackage{amssymb,amsmath}

\begin{document}


\title{Lyapunov Decoherence Rate in Classically Chaotic Systems}


\author{Marcus V. S. Bonan\c{c}a}
\email[]{marcus.bonanca@ufabc.edu.br}
\affiliation{Institut f\"ur Theoretische Physik, Universit\"at Regensburg, D-93040 Regensburg, Germany\\
Centro de Ci\^encias Naturais e Humanas, Universidade Federal do ABC, 09210-170, Santo Andr\'e, SP, Brazil}
\altaffiliation{permanent address}


\date{\today}

\begin{abstract}
We provide a path integral treatment of the decoherence process induced by a heat bath on a single particle 
whose dynamics is classically chaotic and show that the decoherence rate is given by the Lyapunov exponent. 
The loss of coherence is charaterized by the purity, which is calculated semiclassically within diagonal 
approximation, when the particle initial state is a single gaussian wave packet. The calculation is performed 
for weak dissipation and in the high temperature limit. This situation allows us to simplify the heat bath 
description to a single random potential. Although the dissipative term is neglected in such approach, the 
fluctuating one can be treated phenomenologically to fit with the above regime. Our results are therefore valid 
for times shorter than the inverse of the dissipation rate.
\end{abstract}


\maketitle


\section{Introduction}
A well-established way of making classical features to appear in quantum systems is to study them when 
the actions involved are much greater than Planck's constant. This semiclassical approach has allowed us 
to explore the quantum-classical transition in a very interesting class of systems namely those which are 
classically chaotic. In different contexts, several results have shown how the classically chaotic dynamics has 
an influence on the quantum behavior \cite{gutzwiller}. However, it has been shown experimentally that strictly 
quantum features such as superpositions between macroscopically different states are still present even when 
actions are bigger than $\hbar$. In Ref.\cite{wal}, interference patterns measured in Josephson junctions show 
the presence of superpositions between current states, each of those corresponding to the motion of millions of 
Cooper pairs. Therefore, besides the semiclassical regime, a decoherence mechanism must be responsible for the 
emergence of pure classical behavior where such superpositions are absent \cite{zurek1}. 

The decoherence process is one of the effects induced by the coupling to an environment and the path integral 
description of linear systems interacting with a heat bath is very well established \cite{caldeira2,grabert,weiss,breuer}. 
Nevertheless, for nonlinear systems there is no general path integral approach to such problem and specific kinds 
of approximations have been developed \cite{weiss}. At first glance, it can be argued that decoherence happens so 
fast that does not matter whether the dynamics is linear or not \cite{haake}. On the other hand, calculating a 
decoherence rate in such regime contains the assumption that the isolated dynamics of the system of interest is not 
relevant at all. That is true only if the coupling between system and heat bath is weak \cite{breuer} and this is not 
the regime we are interested in here. Once the nonlinearity is relevant, different semiclassical approaches are 
available \cite{heller,grossmann,ozorio1} either within the path integral or the master equation framework. 

In this context, a very interesting result was conjectured for the first time in Ref.\cite{zurek2}. There, the authors 
proposed theoretically that a classically chaotic system could lose coherence at a rate given by the Lyapunov exponent, 
i.e. a rate completely independent of the environment parameters and related to the classically chaotic dynamics. A 
series of works appeared afterwards trying to implement that proposal in solid basis \cite{zurek3,pattanayak,gong,
monteoliva,toscano} (focusing also on the quantum-classical correspondence aspect of the problem). A first approach was 
developed based on the resolution of a master equation for the Wigner function of the classically chaotic system. It was 
shown analytically \cite{pattanayak} and numerically \cite{monteoliva} that the entropy production rate obtained from 
that approach is proportional to the Lyapunov exponent. The analytical calculation was however obtained under the 
following approximations: the chaotic dynamics is taken into account up to its linear regime, the friction from the 
heat bath and the quantum corrections to the classical evolution of the Wigner function are neglected. Despite of this, 
such results were suported by more rigourous calculations considering a hyperbolic linear system \cite{ozorio2}.

A second approach was developed using path integral methods and it is known in the literature by Loschmidt echo (LE)
\cite{jalabert,prosen,jacquod}. The LE is the probability to recover a certain initial state when the Hamiltonian used in the 
forward time evolution is slightly perturbed for the time reversed propagation of the final state. Considered as an 
indirect measure of decoherence \cite{cucchietti1,cucchietti2,casabone,raviola}, the LE was shown to decay with the 
Lyapunov exponent when the system is classically chaotic. On the other hand, the LE deals with isolated systems only 
and therefore it has been understood as a stability measure of them, as proposed originally by Peres \cite{peres}. Thus, 
a path integral formulation of a classically chaotic system interacting with a heat bath still is a very challenging 
problem.

The main goal of the present work is to provide a path integral derivation of the result conjectured in \cite{zurek2} 
avoiding the LE problems mentioned above. In Sec. II the treatment of the heat bath is described and the regime considered 
is presented. In Sec. III we consider the purity as a measure of decoherence and it is calculated through semiclassical 
methods for arbitrary initial states. In Sec. IV it is shown the role played by correlations between pairs of classical 
trajectories in the decay of the purity and how they give rise to a decoherence rate given by the Lyapunov exponent when 
the initial state is a gaussian wave packet of arbitrary width. Finally, conclusions are presented in Sec. V. In the 
appendix we present a brief comparision between LE and purity calculations pointing out their differences.

\section{Description of the Heat Bath}

We consider here a single particle whose dynamics is classically chaotic as, for example, electrons in a clean 
ballistic conductor \cite{richter}. Our goal is to describe its dynamics when it is coupled to a heat bath, 
focusing on the decoherence process. We start from the path integral formulation in terms of Feynman-Vernon 
influence functionals \cite{feynman1} with a Lagrangian given by $L=L_{S}+L_{I}+L_{B}$. The $L_{S}$ term describes 
the isolated dynamics of the particle which we assume to be a classically chaotic billiard-type of confinement. 
The terms $L_{I}$ and $L_{B}$, interaction and heat bath Lagrangians respectively, are assumed to be given by 
Caldeira-Leggett model \cite{caldeira2} with a linear coupling between particle and bath coordinates. For initial 
states of the form $\rho=\rho_{S}\otimes\rho_{B}$, where $\rho_{S}$ and $\rho_{B}$ are the initial density matrices 
of the particle and the bath respectively, the time evolution of the reduced density matrix $\rho_{S}$ is given 
by \cite{feynman1}
\begin{eqnarray}
\rho_{S}(\mathbf{x}_{f},\mathbf{y}_{f};t)=
\int d\mathbf{x}_{i}d\mathbf{y}_{i}J(\mathbf{x}_{f},\mathbf{y}_{f},\mathbf{x}_{i},\mathbf{y}_{i};t)
\rho_{S}(\mathbf{x}_{i},\mathbf{y}_{i};0),\nonumber\\ \label{eq.1}
\end{eqnarray}
where the propagator $J$, obtained after tracing out the heat bath degrees of freedom with $\rho_{B}$ chosen as 
a thermal state, is given by the following path integral
\begin{eqnarray}
 J(\mathbf{x}_{f},\mathbf{y}_{f},\mathbf{x}_{i},\mathbf{y}_{i};t)=\int \mathcal{D}\mathbf{x}\,\mathcal{D}\mathbf{y}\,
\mathrm{e}^{\frac{i}{\hbar}\tilde{S}_{eff}[\mathbf{x},\mathbf{y}]},
\label{eq.2}
\end{eqnarray}

For an Ohmic spectral density and in the high temperature limit, $\tilde{S}_{eff}$ is given by \cite{caldeira2,grabert,weiss}
\begin{equation}
\begin{split}
\tilde{S}_{eff}[\mathbf{x},&\mathbf{y}]= \\
&\tilde{S}_{o}[\mathbf{x},\mathbf{y}]+m\gamma\Phi_{d}[\mathbf{x},\mathbf{y}]
+i\frac{4m\gamma k_{B}T}{\hbar}\Phi_{f}[\mathbf{x},\mathbf{y}],
\end{split}
\label{eq.4}
\end{equation}
$\tilde{S}_{o}[\mathbf{x},\mathbf{y}]=S_{o}[\mathbf{x}]-S_{o}[\mathbf{y}]$, 
$S_{o}[\mathbf{q}]=\int_{0}^{t}ds\,L_{S}[\mathbf{q}(s),\dot{\mathbf{q}}(s)]$ gives the isolated particle dynamics and
the other two terms, $\Phi_{d}$ and $\Phi_{f}$, lead to dissipation and decoherence respectively. They are local quadratic 
functionals of the paths $\mathbf{x}$ and $\mathbf{y}$ without any other prefactors involving the particle mass $m$, 
the friction constant $\gamma$, $\hbar$ or the product $k_{B}T$ of the Boltzmann constant and the heat bath temperature 
\cite{caldeira2,grabert,weiss}. In a regime of very weak dissipation and sufficiently high temperature, the term 
$\Phi_{d}$ is negligible at least for time scales much shorter than $1/\gamma$. In this situation, the effective 
influence of the heat bath can be modeled in a much simpler way.  

Let us consider for a moment the following Lagrangian $L=L_{S}-\mathbf{q}\cdot\mathbf{f}(t)$ where $\mathbf{q}$ is the 
position vector of the system $L_{S}$ and $\mathbf{f}(t)$ is a stochastic force. One can also construct the time 
evolution of the reduced density matrix $\rho_{S}$ in this case but it would depend on $\mathbf{f}(t)$. Thus, assuming 
that $\mathbf{f}(t)$ is gaussian distributed with an average value $\langle \mathbf{f}(t)\rangle=0$, we can calculate 
the following averaged propagator \cite{feynman2}
\begin{eqnarray}
\mathcal{J}(\mathbf{x}_{f},\mathbf{y}_{f},\mathbf{x}_{i},\mathbf{y}_{i};t)=
\int \mathcal{D}\mathbf{f}P[\mathbf{f}]\int \mathcal{D}\mathbf{x}\mathcal{D}\mathbf{y}\,
\mathrm{e}^{\frac{i}{\hbar}\tilde{S}[\mathbf{x},\mathbf{y};\mathbf{f}]},
\label{eq.5}
\end{eqnarray}
where $P[\mathbf{f}]$ is a gaussian functional and
\begin{eqnarray}
 \tilde{S}[\mathbf{x},\mathbf{y};\mathbf{f}]=\tilde{S}_{o}[\mathbf{x},\mathbf{y}]-\int_{0}^{t}ds\,\mathbf{f}(s)\cdot
\Bigl(\mathbf{x}(s)-\mathbf{y}(s)\Bigr).
\label{eq.6}
\end{eqnarray}

The average over $\mathbf{f}(t)$ leads to the effective action 
$\tilde{\mathcal{S}}_{eff}=\tilde{S}_{o}[\mathbf{x},\mathbf{y}]+i\tilde{\mathcal{S}}_{f}[\mathbf{x},\mathbf{y}]$ in 
which \cite{feynman2}
\begin{equation}
\begin{split}
\tilde{\mathcal{S}}_{f}&[\mathbf{x},\mathbf{y}]= \\
&\int_{0}^{t}ds\int_{0}^{s}ds^{\prime}\,\Bigl(\mathbf{x}(s)-\mathbf{y}(s)\Bigr)
\mathbf{\bar{M}}(s-s^{\prime}) \Bigl(\mathbf{x}(s^{\prime})-\mathbf{y}(s^{\prime})\Bigr)
\end{split}
\label{eq.7}
\end{equation}
and $\mathbf{\bar{M}}(s-s^{\prime})$ is $2\times 2$ matrix (since we are considering a two dimensional system) whose 
elements are the correlation functions of the stochastic force
\begin{eqnarray}
 \bar{M}_{i,j}(s-s^{\prime})=\frac{1}{\hbar}\langle f_{i}(s)f_{j}(s^{\prime})\rangle,
\label{eq.8}
\end{eqnarray}
with $f_{i}(t)$, $i=1,2$, being the $\mathbf{f}(t)$ components. Since we are free to choose the spectra of 
$\langle f_{i}(s)f_{j}(\tau)\rangle$, we take it as 
\begin{eqnarray}
 \frac{1}{\hbar}\langle f_{i}(s)f_{j}(s^{\prime})\rangle=\frac{4m\gamma k_{B}T}{\hbar}\delta_{ij}\delta{(s-s^{\prime})}.
\label{eq.9}
\end{eqnarray}
where $\delta_{ij}$ is the Kronecker delta. This choice leads to
\begin{eqnarray}
 \tilde{\mathcal{S}}_{eff}[\mathbf{x},\mathbf{y}]=\tilde{S}_{o}[\mathbf{x},\mathbf{y}]+
i\frac{4m\gamma k_{B}T}{\hbar}\Phi_{f}[\mathbf{x},\mathbf{y}].\label{eq.10}
\end{eqnarray}

Eq.(\ref{eq.10}) is identical to (\ref{eq.4}) except for the dissipative term. Therefore, we can model the influence 
of the heat bath in the regime of very weak dissipation and high temperature just by the fluctuating time dependent 
potential shown in (\ref{eq.6}) and by replacing the propagator (\ref{eq.2}) by the one in (\ref{eq.5}). 

\section{Semiclassical Purity}
The next step then is to write a semiclassical expression for (\ref{eq.5}), which can be written in terms of 
Feynman's propagators
\begin{eqnarray}
\mathcal{J}(\mathbf{x}_{f},&\mathbf{y}_{f}&,\mathbf{x}_{i},\mathbf{y}_{i};t)= \nonumber\\
& &\int \mathcal{D}\mathbf{f}P[\mathbf{f}]K(\mathbf{x}_{f},\mathbf{x}_{i},\mathbf{f};t)K^{*}(\mathbf{y}_{f},
\mathbf{y}_{i},\mathbf{f};t)
\label{eq.11}
\end{eqnarray}
where 
$K(\mathbf{x}_{f},\mathbf{x}_{i},\mathbf{f};t)=\int \mathcal{D}\mathbf{x}\,\mathrm{e}^{iS[\mathbf{x},\mathbf{f}]/\hbar}$, 
$K^{*}$ is its complex conjugate and $S[\mathbf{x},\mathbf{f}]=S_{o}[\mathbf{x}]-\int_{0}^{t}ds\,\mathbf{f}(s)
\cdot\mathbf{x}(s)$.

We assume that the coupling to $\mathbf{f}$ is classically small in order that only the actions, i.e. the phases, 
are affected while the trajectories given by $L_{S}$ remain unchanged. Under this assumption, we replace $K$ by 
the semiclassical Van Vleck formula \cite{gutzwiller} for two dimensional systems
\begin{eqnarray}
K_{sc}&&(\mathbf{x}_{f},\mathbf{x}_{i},\mathbf{f};t)= \nonumber\\
\Big(&&\frac{1}{2\pi i\hbar}\Big)\sum_{\tilde{\alpha}(\mathbf{x}_{i}\rightarrow\mathbf{x}_{f},t)}
D_{\tilde{\alpha}}\exp{\left(\frac{i}{\hbar}S_{\tilde{\alpha}}(\mathbf{x}_{f},\mathbf{x}_{i},\mathbf{f};t)\right)}
\label{eq.12}
\end{eqnarray}
where $S_{\tilde{\alpha}}$ is the classical action $S$ of the trajectory $\tilde{\alpha}$ running from $\mathbf{x}_{i}$ 
to $\mathbf{x}_{f}$ in time $t$, and $D_{\tilde{\alpha}}=|\mathrm{det}(\partial^{2}S_{\tilde{\alpha}}/
\partial\mathbf{x}_{i}\partial\mathbf{x}_{f})|^{1/2}
\exp{(-\frac{i\pi}{2}\mu_{\tilde{\alpha}})}$ is the Van Vleck determinant including the Maslov index.

To characterize the decoherence process, we calculate the purity $\mathrm{Tr}\left(\rho^{2}_{S}(t)\right)$, where 
the trace is performed over the particle degrees of freedom since $\rho_{S}$ is already a reduced density matrix. 
If $\rho_{S}$ is initially pure, the purity starts from one and decays as time evolves due to heat bath influence. 
The semiclassical expression for the purity is therefore given by the trace of the product of two $\rho_{S}(t)$ each 
one evolved by the semiclassical version of $\mathcal{J}$ calculated from (\ref{eq.11}) when $K$ is replaced by 
(\ref{eq.12}). That yields the following expression which recalls the LE one \cite{cucchietti2,gutkin}
(we refer to Sec. V for a brief discussion about this issue)
\begin{widetext}
\begin{eqnarray}\label{eq.13}
\mathrm{Tr}\left(\rho^{2}_{S}(t)\right)&=&
\left(\frac{1}{2\pi\hbar}\right)^{4}\int d\mathbf{x}_{f}d\mathbf{y}_{f} d\mathbf{x}_{i}d\mathbf{y}_{i} d\mathbf{x}^{\prime}_{i}d\mathbf{y}^{\prime}_{i}\int D\mathbf{f}D\mathbf{g}\,P[\mathbf{f}]\,P[\mathbf{g}]\,
\rho_{S}(\mathbf{x}_{i},\mathbf{y}_{i};0)\,\rho_{S}(\mathbf{y}^{\prime}_{i},\mathbf{x}^{\prime}_{i};0) \nonumber\\
& &\times
\sum_{\substack{\tilde{\alpha}(\mathbf{x}_{i}\rightarrow\mathbf{x}_{f},t), \\  \tilde{\alpha}^{\prime}(\mathbf{y}_{i}\rightarrow\mathbf{y}_{f},t)}}\;
\sum_{\substack{\tilde{\eta}(\mathbf{y}^{\prime}_{i}\rightarrow\mathbf{y}_{f},t), \\
\tilde{\eta}^{\prime}(\mathbf{x}^{\prime}_{i}\rightarrow\mathbf{x}_{f},t)}}
D_{\tilde{\alpha}}D^{*}_{\tilde{\alpha}^{\prime}}D_{\tilde{\eta}}D^{*}_{\tilde{\eta}^{\prime}}\\
& &\times\exp{\left[\frac{i}{\hbar}\Bigl(
S_{\tilde{\alpha}}(\mathbf{x}_{f},\mathbf{x}_{i},\mathbf{f};t)-
S_{\tilde{\alpha}^{\prime}}(\mathbf{y}_{f},\mathbf{y}_{i},\mathbf{f};t)+
S_{\tilde{\eta}}(\mathbf{y}_{f},\mathbf{y}^{\prime}_{i},\mathbf{g};t)-
S_{\tilde{\eta}^{\prime}}(\mathbf{x}_{f},\mathbf{x}^{\prime}_{i},\mathbf{g};t)
\Bigr)\right]}\nonumber
\end{eqnarray}
\end{widetext}

Due to the rapidly oscillatory phase factor containing the action differences, most of the contributions will 
cancel out except for the semiclassically small action differences originating from pairs of trajectories which are 
close to each other in configuration space. We can thus use a linear approximation in order to relate the actions 
$S_{\tilde{\alpha}}$, $S_{\tilde{\eta}^{\prime}}$ along the trajectories $\tilde{\alpha},\,\tilde{\eta}^{\prime}$ to
 the actions $S_{\alpha},\,S_{\eta^{\prime}}$ along nearby trajectories $\alpha,\,\eta^{\prime}$ connecting the midpoint 
$\mathbf{r}_{i}=(\mathbf{x}_{i}+\mathbf{x}^{\prime}_{i})/2$ with $\mathbf{x}_{f}$. In the same way we relate 
$S_{\tilde{\alpha}^{\prime}},\,S_{\tilde{\eta}}$ to the actions $S_{\alpha^{\prime}},\,S_{\eta}$ along the nearby 
trajectories $\alpha^{\prime},\,\eta$ connecting the midpoint $\mathbf{r}^{\prime}_{i}=
(\mathbf{y}_{i}+\mathbf{y}^{\prime}_{i})/2$ with $\mathbf{y}_{f}$. The expansion will contain terms up to zero 
order in the $D$'s and terms up to first order in the exponential. For $S_{\tilde{\alpha}}$ and 
$S_{\tilde{\alpha}^{\prime}}$, the linearization yields \cite{gutkin}
\begin{equation}
\begin{split}
S_{\tilde{\alpha}}(\mathbf{x}_{f},\mathbf{x}_{i},&\mathbf{f};t)\approx  \\
 & S_{o,\alpha}(\mathbf{x}_{f},\mathbf{r}_{i};t)-
\int_{0}^{t}ds\,\mathbf{q}_{\alpha}(s)\cdot\mathbf{f}(s)-\frac{1}{2}\mathbf{u}\cdot\mathbf{p}_{i}^{\alpha}, \\
S_{\tilde{\eta}}(\mathbf{y}_{f},\mathbf{y}^{\prime}_{i},&\mathbf{g};t)\approx \\
 & S_{o,\eta}(\mathbf{y}_{f},\mathbf{r}^{\prime}_{i};t)-
\int_{0}^{t}ds\,\mathbf{q}_{\eta}(s)\cdot\mathbf{g}(s)+\frac{1}{2}\mathbf{u}^{\prime}\cdot\mathbf{p}_{i}^{\eta}
\end{split}
\label{eq.14}
\end{equation}
where $S_{o,\alpha}$ is $S_{o}[\mathbf{x}]$ along the trajectory $\alpha$, $\mathbf{u}=
(\mathbf{x}_{i}-\mathbf{x}^{\prime}_{i})$, $\mathbf{u}^{\prime}=(\mathbf{y}_{i}-\mathbf{y}^{\prime}_{i})$ and 
$\mathbf{p}^{\alpha}_{i},\,\mathbf{p}^{\eta}_{i}$ are the initial momenta of trajectories $\alpha,\,\eta$. 
Analagous expressions are obtained for $S_{\tilde{\alpha}^{\prime}}$ and $S_{\tilde{\eta}^{\prime}}$. As mentioned 
before, for the prefactors we have $D_{\tilde{\alpha}}\approx D_{\alpha}$, $D_{\tilde{\eta}}\approx D_{\eta}$ and 
analogously for $D_{\tilde{\alpha}^{\prime}}$, $D_{\tilde{\eta}^{\prime}}$.

In order to evaluate the sums over paths in (\ref{eq.13}) we consider only the diagonal contribution obtained by 
the pairings $\alpha=\eta^{\prime}$ and $\alpha^{\prime}=\eta$. Besides that, we perform the gaussian averages 
over $\mathbf{f}$ and $\mathbf{g}$ as described in (\ref{eq.7}). Using (\ref{eq.8}) and (\ref{eq.9}) we obtain
\begin{eqnarray}\label{eq.15}
 \mathrm{Tr}\left(\rho^{2}_{S}(t)\right)&=&\int d\mathbf{x}_{f}d\mathbf{y}_{f}d\mathbf{r}_{i}d\mathbf{r}^{\prime}_{i}
\rho^{W}_{S}(\mathbf{r}_{i},\mathbf{p}^{\eta}_{i})\rho^{W}_{S}(\mathbf{r}^{\prime}_{i},\mathbf{p}^{\alpha}_{i})
\nonumber\\
&\times&\sum_{\substack{\alpha(\mathbf{r}_{i}\rightarrow\mathbf{x}_{f},t),\\ 
\eta(\mathbf{r}^{\prime}_{i}\rightarrow\mathbf{y}_{f},t)}}
|D_{\alpha}|^{2}|D_{\eta}|^{2} \\
&\times&\exp{\left[-\frac{2\kappa}{\hbar^{2}}\int_{0}^{t} ds\Bigl(\mathbf{q}_{\eta}(s)-
\mathbf{q}_{\alpha}(s)\Bigr)^{2}\right]}
\nonumber
\end{eqnarray}
where $\kappa=4m\gamma k_{B}T$ and
\begin{eqnarray}
\rho^{W}_{S}(\mathbf{r},\mathbf{p})=
\left(\frac{1}{2\pi\hbar}\right)^{2}\int d\mathbf{u}\, \rho_{S}\left(\mathbf{r}+\frac{\mathbf{u}}{2},\mathbf{r}-
\frac{\mathbf{u}}{2};0\right)\,
 \mathrm{e}^{-\frac{i}{\hbar}\mathbf{u}\cdot\mathbf{p}}\nonumber\\
\label{eq.16}
\end{eqnarray}
is the Wigner function of the initial $\rho_{S}$. 

\section{Decoherence rates}
We consider two kinds of contributions in order to evaluate (\ref{eq.15}). First, if $\alpha$ and $\eta$ are close 
enough to each other in phase space, it is possible to linearize the motion of one trajectory around the other to obtain
\begin{eqnarray}
\mathbf{q}_{\eta}(s)-\mathbf{q}_{\alpha}(s)&\approx& \left((\mathbf{r}^{\prime}_{i}-\mathbf{r}_{i})+\frac{1}{m\lambda}
(\mathbf{p}^{\eta}_{i}-\mathbf{p}^{\alpha}_{i})\right)
\frac{\mathrm{e}^{\lambda s}}{2} \nonumber\\
&+&\left((\mathbf{r}^{\prime}_{i}-\mathbf{r}_{i})-\frac{1}{m\lambda}(\mathbf{p}^{\eta}_{i}-\mathbf{p}^{\alpha}_{i})\right)
\frac{\mathrm{e}^{-\lambda s}}{2} \nonumber\\
\label{eq.17}
\end{eqnarray}
where $\lambda$ is the Lyapunov exponent. On the other hand, if they are not close $\alpha$ and $\eta$ can be first 
considered as free particle trajectories for time scales shorter than an average free flight time $t_{o}$. In this case,
\begin{eqnarray}
 \mathbf{q}_{\eta}(s)-\mathbf{q}_{\alpha}(s)\approx
 (\mathbf{r}^{\prime}_{i}-\mathbf{r}_{i})+\frac{s}{m}(\mathbf{p}^{\eta}_{i}-\mathbf{p}^{\alpha}_{i}).
\label{eq.18}
\end{eqnarray}
After $t_{o}$, we can only estimate $\left(\mathbf{q}_{\eta}(s)-\mathbf{q}_{\alpha}(s)\right)^{2}$ by its average 
value over the billiard area $A$. Assuming ergodicity, this average can be calculated as
\begin{equation}
 \left\langle \left(\mathbf{q}_{\eta}(s)-\mathbf{q}_{\alpha}(s)\right)^{2}\right\rangle_{A}=\int_{A}
 d\mathbf{q}\,d\mathbf{q}^{\prime}\frac{(\mathbf{q}-\mathbf{q}^{\prime})^{2}}{A^{2}}. \label{eq.19}
\end{equation}
It can be verified for simple geometries that (\ref{eq.19}) yields $\langle \left(\mathbf{q}_{\eta}(s)-
\mathbf{q}_{\alpha}(s)\right)^{2}\rangle_{A}\propto A$. 

The prefactors $|D_{\alpha}|^{2}$ and $|D_{\eta}|^{2}$ in (\ref{eq.15}) can be regarded as Jacobians when
 transforming the integrals over the final positions $\mathbf{x}_{f}$ and $\mathbf{y}_{f}$ into integrals over 
the initial momenta $\mathbf{p}^{\eta}_{i}\equiv \mathbf{p}_{i}$ and $\mathbf{p}^{\alpha}_{i}\equiv 
\mathbf{p}^{\prime}_{i}$ \cite{jalabert,jacquod} leading to
\begin{eqnarray}
\mathrm{Tr}\left(\rho^{2}_{S}(t)\right)&=&\int d\mathbf{r}_{i}d\mathbf{r}^{\prime}_{i}d\mathbf{p}_{i}d\mathbf{p}^{\prime}_{i}
\rho^{W}_{S}(\mathbf{r}_{i},\mathbf{p}_{i})\rho^{W}_{S}(\mathbf{r}^{\prime}_{i},\mathbf{p}^{\prime}_{i})
\nonumber\\
&\times&\exp{\left[-\frac{2\kappa}{\hbar^{2}}\int_{0}^{t} ds\, 
Q^{2}(\mathbf{r}_{i},\mathbf{r}^{\prime}_{i},\mathbf{p}_{i},\mathbf{p}^{\prime}_{i};s)\right]}
\label{20.1}
\end{eqnarray}
where the function $Q^{2}$ is taken either by the square of (\ref{eq.17}) or (\ref{eq.18}) or simply by (\ref{eq.19}) 
replacing $\mathbf{p}^{\eta}_{i}$ by $\mathbf{p}_{i}$ and  $\mathbf{p}^{\alpha}_{i}$ by $\mathbf{p}^{\prime}_{i}$. 
Thus, for initial states given by single gaussian wave packets as
\begin{eqnarray}
\lefteqn{
\rho_{S}\left(\mathbf{r}+\frac{\mathbf{u}}{2},\mathbf{r}-\frac{\mathbf{u}}{2};0\right)=}\nonumber \\
& &\frac{1}{\left(\pi\sigma^{2}\right)}\exp{\left(-\frac{(\mathbf{r}-\mathbf{r}_{o})^{2}}{\sigma^{2}}-
\frac{\mathbf{u}^{2}}{4\sigma^{2}}+\frac{i}{\hbar}\mathbf{p}_{o}\cdot\mathbf{u}\right)},
\label{eq.20}
\end{eqnarray}
$\mathrm{Tr}\left(\rho^{2}_{S}(t)\right)$ can be calculated for the different situations mentioned above.

Defining $\tau\equiv\gamma t$ and using (\ref{eq.18}), the result for times between $0$ and $\tau_{o}\equiv\gamma t_{o}$ 
is 
\begin{eqnarray}
 \mathrm{Tr}\left(\rho^{2}_{S}(\tau)\right)=\left[1+16a_{1}\tau+\frac{2}{3}a_{2}\tau^{3}\left(1+
\frac{a_{1}}{2}\tau\right)\right]^{-1}
\label{eq.21}
\end{eqnarray}
where $a_{1}=k_{B}T/\bar{E}$, $\bar{E}=\frac{\hbar^{2}}{2m\sigma^{2}}$ and $a_{2}=D/\gamma\sigma^{2}$ with 
$D=\frac{4k_{B}T}{m\gamma}$. Since our model is valid for times much shorter than $1/\gamma$, $\tau$ is certainly 
smaller than 1. The values of $a_{1}$ and $a_{2}$ define relations between the free parameters of the problem. It 
is important to mentioned that we have not assumed at any point a highly localized initial state. Therefore, 
$a_{2}\sim 1$ when $\sigma^{2}$ is comparable to $D/\gamma$ and $a_{2}\gg 1$ for localized wave packets. 
Eq. (\ref{eq.21}) shows that the purity starts from one, as it should be since the initial state (\ref{eq.20}) is 
pure, and decreases as $\tau$ approaches $\tau_{o}$.

When $\tau>\tau_{o}$, there are two contributions: either the trajectories are close to each other in the sense 
mentioned before or they are not. In the latter case, since (\ref{eq.19}) is mainly $A$, (\ref{eq.15}) together 
with (\ref{eq.20}) yield
\begin{eqnarray}
 \mathrm{Tr}\left(\rho^{2}_{S}(\tau)\right)=\exp{\left(-16\pi\frac{k_{B}T}{\Delta}(\tau-\tau_{o})\right)},\label{eq.22}
\end{eqnarray}
where $\Delta=\frac{2\pi\hbar^{2}}{mA}$ is the mean level spacing for a billiard of area $A$. In the regime considered 
here, $k_{B}T\gg\Delta$ and (\ref{eq.22}) decays very fast.

When the trajectories are correlated by the Lyapunov spreading, (\ref{eq.15}), (\ref{eq.17}) and (\ref{eq.20}) yield
\begin{eqnarray}
 \mathrm{Tr}\left(\rho^{2}_{S}(\tau)\right)&=&
\biggl\{1+b_{1}\left[(1+b_{2})\sinh{(2\Lambda\tau)}+(1-b_{2})2\Lambda\tau\right]\nonumber \\
&+&b_{3}\left[\cosh{(2\Lambda\tau)}-\frac{(2\Lambda\tau)^{2}}{2}-1\right]\biggr\}^{-1},
\label{eq.23}
\end{eqnarray}
where $b_{1}=\frac{1}{\Lambda}\frac{k_{B}T}{\bar{E}}$, $b_{2}=(2\bar{E}/\hbar\lambda)^{2}$, 
$b_{3}=32\left(\frac{1}{\Lambda}\frac{k_{B}T}{\hbar\lambda}\right)^{2}$ and $\Lambda=\lambda/\gamma$. 
For $\Lambda \tau>1$, only those terms with $\mathrm{e}^{2\Lambda\tau}$ are relevant and (\ref{eq.23}) can be written as
\begin{eqnarray}
 \mathrm{Tr}\left(\rho^{2}_{S}(\tau)\right)\approx\left(1+\beta\mathrm{e}^{2\Lambda\tau}\right)^{-1},
\label{eq.24}
\end{eqnarray}
where $\beta=b_{1}\frac{(1+b_{2})}{2}+\frac{b_{3}}{2}$. Eq. (\ref{eq.23}) shows that the classically chaotic 
dynamics induces a decoherence rate independent of the heat bath parameters.

\section{Conclusions and Discussions}
Summarizing, we have shown how to describe the decoherence process of a classically chaotic system coupled to 
a heat bath within a path integral framework. Our approach leads to a decoherence rate given by the Lyapunov 
exponent. Although the results above were derived only for single gaussian wave packets with arbitrary widths as 
initial states, the semiclassical approach presented also allows the treatment of their superpositions. This is an 
important difference compared to previous works where only localized wave packets have been considered 
\cite{zurek2,monteoliva,jalabert,cucchietti2}. The results for superpositions will be presented elsewhere 
and will be compared to the predictions of Refs.\cite{pattanayak,ozorio2} which have claimed that the Lyapunov regime 
does not depend on the initial state. Concerning the description of the stable and unstable directions in (\ref{eq.17}), 
that is not the most general one but it certainly captures the qualitative behavior one should find in a specific case. 
Different time scales of the classical chaotic dynamics were taken into account in the present calculation differently 
from those in Refs.\cite{pattanayak,ozorio2} where it is considered up to its linear regime only. 

It is possible to extend the present results beyond the high-temperature limit as long as the correlation time of 
thermal fluctuations is much shorter than the time scales of the system of interest dynamics \cite{breuer}. Corrections 
to the diagonal approximation performed here could also be studied \cite{richter} for action differences of the order 
of $\hbar$. As in the LE case \cite{gutkin}, they would take quantum effects into account systematically. Dissipation 
was neglected in the present calculation making the heat bath treatment comparable to previous ones using 
master equations \cite{pattanayak,monteoliva}. However when the coupling to the heat bath is strong enough, 
dissipative terms cannot be neglected and it is still an open question whether the Lyapunov decoherence rate would be 
robust to that.

Concerning the experimental observation of our theoretical results, it was predicted recently in Ref.\cite{pozzo} 
that Josephson junctions devices could also be used to observe the Lyapunov regime in the time evolution of the 
fidelity. However, those devices are almost isolated from the heat bath in that context. We believe that the same 
regime and initial state preparation described there could be used to study the decay of interference fringes and 
to observe the Lyapunov decoherence rate as long as the heat bath temperature is increased. 

We briefly compare now the semiclassical calculation of the purity presented here and those of the LE 
\cite{jalabert,cucchietti2,gutkin}. At first glance, eq.(\ref{eq.13}) is identical to eq.(5.7) in Ref.\cite{cucchietti2} or 
to eq.(73) in Ref.\cite{gutkin}. They certainly have one thing in common: all of them are given in terms of four sums 
over chaotic trajectories which are considered unperturbed either by the heat bath influence (in the purity case) or by 
the extra potential (in the LE case). In other words, those equations result from the same perturbative approach that 
allows us to use the well-known properties of those trajectories. One obtains then from those equations the leading order 
result after performing diagonal approximation, which is another common point they have. 

The first subtle (though important) difference concerns the initial state. In Ref.\cite{cucchietti2} for example, as 
most of the semiclassical analytical works on LE, a further approximation is performed over eq.(5.7) since a highly 
localized gaussian wave packet is considered as initial state leading to eq.(5.11). A more general calculation however 
was done recently in Ref.\cite{gutkin} allowing any kind of initial state and which we have followed closely here. 
The second difference concerns the averages. Eq.(73) in Ref.\cite{gutkin} shows very clearly that LE is mostly calculated 
as an average of fidelity amplitude squared modulus. On the other hand, the purity is obtained from the trace of the 
reduced density matrix squared, i.e. of the product of two averaged quantities. Hence the averages over the random 
potential that appear in (\ref{eq.13}) are always performed independently. In LE case, the perturbation often depends 
on the trajectories which may be correlated or not. Thus averages cannot be performed always independently. Nevertheless 
both purity and LE semiclassical expressions have two kinds of contributions arising from uncorrelated and correlated 
trajectories, the last one leading to the Lyapunov regime. The uncorrelated trajectories give rise to the Fermi golden 
rule regime in the LE case, which sometimes dominates its decay. For the purity they lead to a very big decoherence 
rate (eq.(\ref{eq.22})) that kills this contribution very fast and makes the Lyapunov rate always the dominant one 
(this result agrees with the numerical ones in Ref.\cite{monteoliva}). Finally, we have shown that the Lyapunov regime 
can be obtained from the exponents (eq.(\ref{eq.16}) and (\ref{20.1})) instead of the prefactors 
\cite{jalabert,cucchietti2} avoiding short time problems.

\begin{acknowledgments}
The author acknowledges support of the Brazilian research agency CNPq and DFG (GRK 638). The author is 
also grateful to M. Guti\'errez, D. Waltner, C. Petitjean, R. A. Jalabert and K. Richter for valuable 
discussions and J. Hausinger and J. D. Urbina for their careful reading of the manuscript and valuable suggestions. 
\end{acknowledgments}


%

\end{document}